\documentclass{jps-cp}
\usepackage{txfonts} %Please comment out this line unless the txfonts package is availabe in your LaTeX system.

\title{Large-Scale Numerical-Diagonalization Study of the Shastry-Sutherland Model}

\author{Hiroki \textsc{Nakano}$^{1}$ and T$\hat{\rm o}$ru \textsc{Sakai}$^{1,2}$}

\inst{$^{1}$Graduate School of Science, University of Hyogo, Kamigori, Hyogo 678-1297, Japan \\
$^{2}$National Institutes for Quantum and Radiological Science and Technology,
SPring-8 Sayo, Hyogo 679-5148, Japan}

\email{hnakano@sci.u-hyogo.ac.jp}

\recdate{July 14, 2022}

\abst{The $S=1/2$ Heisenberg antiferromagnet on the orthogonal-dimer lattice 
(Shastry-Sutherland model) 
is studied by the Lanczos diagonalization method. 
The properties of this model are determined by the ratio of two interactions, 
namely, $r=J_2/J_1$, 
where $J_1$ denotes the amplitude of spin interactions at orthogonal dimers 
and the interactions represented by $J_2$ form the square lattice. 
We focus our attention on the edge of the phase in which 
the dimer state is realized as the exact ground state. 
Our large-scale calculations of diagonalizations 
treating finite-size clusters including 44 and 48 spin sites 
successfully detect the target edge. 
Our conclusion is that the ratio for the edge is $r= 0.6754(2)$. 
This estimate is compared with an experimental result 
from electron spin resonance measurements. 
}

\kword{$S=1/2$ Heisenberg antiferromagnet, orthogonal-dimer lattice, 
Lanczos diagonalization, parallel calculations}
  
\begin{document}
\maketitle

\section{Introduction}

The Shastry-Sutherland model
\---
the $S=1/2$ Heisenberg antiferromagnet on the orthogonal-dimer lattice 
\---
has attracted considerable attention 
as a typical frustrated magnet\cite{ShastrySutherland1982}. 
The most characteristic behavior of this model is that 
the system realizes the ground state that can be expressed 
by the mathematically exact quantum state, namely, the singlet-dimer state, 
when the ratio $J_{2}/J_{1}$ is smaller than a specific value, 
where $J_{1}$ and $J_{2}$ denote the orthogonal-dimer interaction and 
the interaction forming the square lattice, respectively. 
On the other hand, the N$\acute{\rm e}$el-ordered ground state is realized when
the ratio $J_{2}/J_{1}$ is sufficiently large. 
However, the properties of this model are still unclear 
in and near the intermediate range 
between the exact dimer state and the N$\acute{\rm e}$el-ordered state. 
SrCu$_{2}$(BO$_{3}$)$_{2}$ is 
a good candidate material for this model\cite{KageyamaPRL1999}; 
it is considered that this material corresponds to the case when 
the ratio $J_{2}/J_{1}$ is in the exact dimer phase. 
Recent experimental reports have shown 
that the ratio can be changed under pressure 
and 
that one can detect the edge 
of the exact dimer phase\cite{Zayed_NatPhys2017,Sakurai_JPSJ2017}. 
On the other hand,  it is still difficult to precisely determine 
the phase transition points theoretically during the variation of $J_{2}/J_{1}$ 
because of the strong frustration in this model. 
In several numerical studies, 
the estimation of the transition points has been tackled. 
Among numerical-diagonalization studies, the Shastry-Sutherland model has been 
investigated for finite-size clusters with sizes 
up to 40 spin sites\cite{HNakano_JPSJ2018}. 

Let us focus our attention on the phase boundary 
between the exact dimer phase and its neighboring phase. 
This boundary has been experimentally detected as mentioned before. 
If we obtain a theoretical result for the boundary as a precise estimate, 
the experimental and theoretical results can be compared; 
it will contribute much to our understanding of this system. 
Under such circumstances, 
the purpose of this study is to estimate the ratio $J_{2}/J_{1}$ 
for this boundary from additional results of system sizes 
that have not been treated before. 
In this study, we successfully carried out 
Lanczos diagonalization calculations to obtain the ground-state energy 
of finite-size clusters larger than those in previous studies.  
Our calculations enable us to obtain 
a highly precise estimation 
for the transition point for the edge of the exact dimer phase. 
Our additional results show a weak system-size dependence 
of the ratio for the transition point. 
Our estimation for the transition point will be compared
with an experimental observation. 

This paper is organized as follows. 
In the next section, we will introduce the model Hamiltonian 
and explain our numerical method. 
In the third section, 
we will present and discuss our results. 
In the final section, 
we will summarize the results of this study and provide some remarks. 

\section{Hamiltonian and Method}

\begin{figure}[tbh]
\includegraphics[width=15cm]{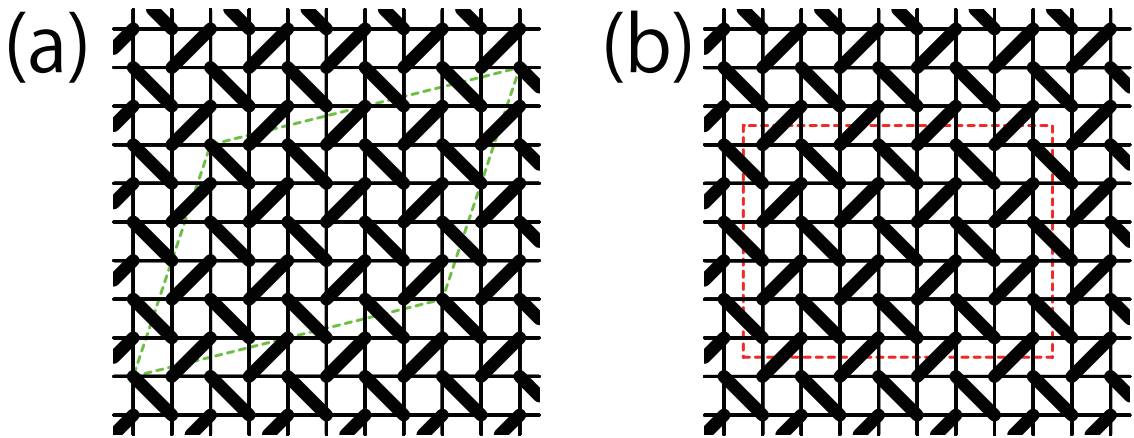}
\caption{Structure of the orthogonal-dimer lattice and its finite-size clusters.
Thick and thin solid lines denote 
bonds for $J_1$ and $J_2$, respectively. 
Panel (a) shows the finite-size cluster of $N_{\rm s}=44$ 
indicated by green dotted lines.
Panel (b) shows the finite-size cluster of $N_{\rm s}=48$ 
indicated by red dotted lines.
}
\label{fig1}
\end{figure}

The Hamiltonian studied here is given by 
%\begin{eqnarray}
\begin{equation}
{\cal H}
%&=& 
=
\sum_{\langle i ,j\rangle : \ {\rm orthogonal~dimer}} J_{1} 
\mbox{\boldmath $S$}_{i}\cdot\mbox{\boldmath $S$}_{j} 
+
\sum_{\langle i ,j\rangle : \ {\rm square~lattice}} J_{2} 
\mbox{\boldmath $S$}_{i}\cdot\mbox{\boldmath $S$}_{j} 
, 
\label{Hamiltonian}
%\end{eqnarray}
\end{equation}
where $\mbox{\boldmath $S$}_{i}$ 
denotes the $S=1/2$ spin operator at site $i$. 
Here, we consider the case of an isotropic interaction in spin space. 
Site $i$ is assumed to characterize the vertex 
of the square lattice. 
The number of spin sites is represented by $N_{\rm s}$. 
The first and second terms of Eq.~(\ref{Hamiltonian}) denote 
orthogonal dimer interactions represented 
by thick solid bonds in Fig.~\ref{fig1} and 
interactions forming the square lattice
represented by thin solid bonds in Fig.~\ref{fig1}, respectively. 
The two interactions between the two spins 
are antiferromagnetic, namely, $J_{1} > 0$ and $J_{2} > 0$. 
Note here that energies are measured in units of $J_{1}$; 
hereafter, we set $J_{1}=1$. 
The ratio $J_{2}/J_{1}$ is denoted by $r$;  
when $r=0$, the system is an assembly of isolated dimerized-spin models;  
on the other hand, the system in the limit $r\rightarrow\infty$ is reduced 
to the $S=1/2$ Heisenberg antiferromagnet on the ordinary square lattice. 
We treat finite-size clusters with system size $N_{\rm s}$ 
under the periodic boundary condition. 
In this study, we additionally treat $N_{\rm s}=44$ and 48;  
finite-size clusters for these cases are shown in Fig.~\ref{fig1}. 
Unfortunately, finite-size clusters for $N_{\rm s}=44$ and 48 
cannot form squares even if the squares are tilted; 
therefore, the treated clusters are not squares. 
Even under this situation, 
calculations for the treated clusters can provide us 
with significant information concerning this system. 

In this study, numerical diagonalizations
are carried out on the basis of the Lanczos algorithm 
to obtain the lowest energy of ${\cal H}$ 
in the subspace characterized by $\sum _j S_j^z=M$. 
The $z$-axis is taken as the quantized axis of each spin.  
Numerical-diagonalization calculations are widely considered to be unbiased. 
Thus, one can obtain reliable information about the target system. 
The energy is represented by $E(N_{\rm s},M)$;  
here, we calculate the energy in the case of $M=0$ 
because our attention is focused primarily 
on the behavior of the ground-state energy
$E_{\rm g}=E(N_{\rm s},0)$. 
Some of the Lanczos diagonalizations were carried out 
using the MPI-parallelized code that was originally 
developed in the study of Haldane gaps\cite{HNakano_JPSJ2009}. 
The usefulness of our program was confirmed in large-scale 
parallelized calculations in various 
studies\cite{HNakano_JPSJ2011,HNakano_JPSJ2018,
HNakano_JPSJ2019,HNakano_S1HaldaneGap_JPSJ2022}. 
Our largest-scale calculations for $N_{\rm s}=48$ in this study 
have been carried out using Fugaku.
The calculations for $N_{\rm s}=48$ in Fugaku use 65536 nodes that
correspond to approximately 41\% of the nodes in Fugaku. 
Note here that 
the dimension of $N_{\rm s}=48$ and $M=0$ is 32,247,603,683,100 
and
that this dimension is larger than 18,252,025,766,941
in the case of $N=30$ for $S=1$
in which Lanczos diagonalization was successfully carried out 
in Ref.~\ref{HNakano_S1HaldaneGap_JPSJ2022}. 

\section{Results and Discussion}

In this study, the ratio of interactions 
for the edge of the exact dimer phase is focused on; 
this ratio is denoted by $r_{\rm c1}$ hereafter. 
Before presenting our results, 
let us review previous estimates of $r_{\rm c1}$. 
Estimates of $r_{\rm c1}$ are summarized 
in chronological order in Table~\ref{table_1}. 
Note here that Refs.~\ref{HNakano_JPSJ2018}, \ref{MiyaharaPRL1999}, 
and \ref{LauchliPRB2002} were based on the method 
of numerical diagonalizations. 
On the other hand,
in Refs.~\ref{Lou_arXiv1212_1999} and \ref{Corboz_PRB2013},  
a type of calculation based on a tensor-network framework was employed. 
In Ref.~\ref{KogaPRL2000}, a series expansion method was used.
Note that Ref.~\ref{KogaPRL2000} was the first report 
that pointed out the existence of an intermediate phase 
between the exact dimer and N$\acute{\rm e}$el-ordered phases. 
\begin{table}[tbh]
\caption{Estimates of the ratio of interactions $r_{\rm c1}$ corresponding 
to the edge of the exact dimer phase from various theoretical approaches. 
} 
\label{table_1}
\begin{tabular}{ccll}
\hline
Ref. & Publication year & $r_{\rm c1}$ & Method \\
\hline
\ref{Albrecht1996}        & 1996    & 0.6       & Schwinger boson mean field theory \\
\ref{MiyaharaPRL1999}     & 1999    & 0.70(1)   & Numerical Diagonalization~(up to 20 sites) \\
\ref{WeihongPRB1999}      & 1999    & 0.691(6)  & Ising Expansion \\
\ref{MullerHartmannPRL2000}&2000    & 0.697(2)  & Dimer Expansion \\
\ref{KogaPRL2000}         & 2000    & 0.677(2)  & Plaquette Expansion \\
\ref{LauchliPRB2002}      & 2002    & 0.678     & Numerical Diagonalization~(up to 32 sites) \\
\ref{Lou_arXiv1212_1999}  & 2012    & 0.687(3)  & Tensor Network with MERA \\
\ref{Corboz_PRB2013}      & 2013    & 0.675(2)  & Tensor Network with iPEPS \\
\ref{HNakano_JPSJ2018}    & 2018    & 0.675     & Numerical Diagonalization~(36 and 40 sites) \\
This study                & present & 0.6754(2) & Numerical Diagonalization~(44 and 48 sites) \\
\hline
\end{tabular}
\end{table}

\begin{figure}[tbh]
\includegraphics[width=15cm]{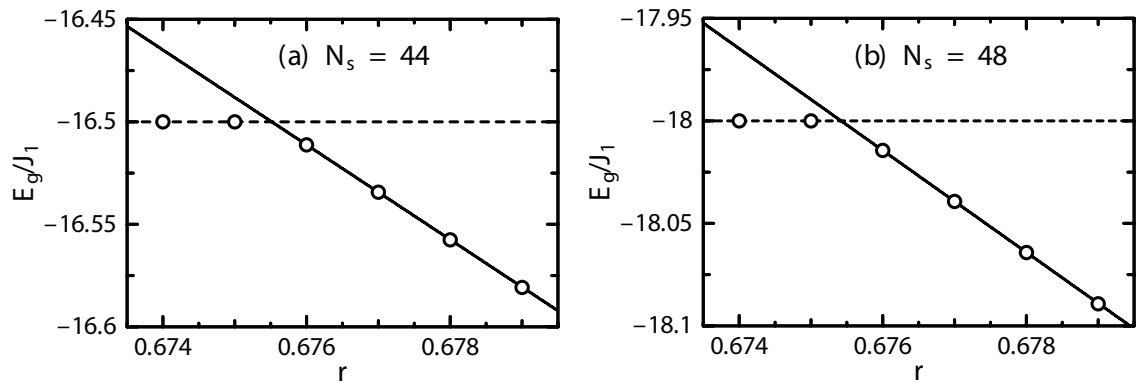}
\caption{Ground state energy for $N_{\rm s}=44$ and 48 
in panels (a) and (b), respectively. 
Circles denote numerical-diagonalization results.
Horizontal broken lines represent the energy level 
of the exact dimer state. 
Solid lines are obtained by the fitting based on the two points 
that are close to the energy level of the exact dimer state. 
}
\label{fig2}
\end{figure}

Now, let us observe the $r$-dependence of the ground state energy 
of the Hamiltonian~(\ref{Hamiltonian}); 
the results are shown in Fig.~\ref{fig2}. 
One finds that 
our calculations for both $N_{\rm s}=44$ and 48 have succeeded 
in capturing the energy level ($E_{\rm g}=-(3/8)N_{\rm s}$)  
of the exact dimer state in the region of small $r$. 
In Fig.~\ref{fig2}, this energy level is shown 
by the broken line for each $N_{\rm s}$. 
In the region where $r$ becomes larger than a specific value, 
the ground state energy at the same time becomes lower than
the energy level of the exact dimer state. 
In Fig.~\ref{fig2}, 
we draw a fitting line determined from the two data points 
that are close to the energy level of the exact dimer state. 
Let us focus our attention on large-$r$ data points 
that are distant from the energy level of the exact dimer state. 
One also confirms that these points fall on the fitting line,  
although these points are not used for the fitting.  
This behavior suggests that 
in the range of large $r$ in Fig.~\ref{fig2}, there appears 
a spin state that differs from the exact dimer state. 
From the point where the horizontal broken line and 
the solid fitting line cross, 
we can obtain the information concerning the edge of the exact dimer phase. 
Therefore, our present results are 
\begin{equation}
r_{\rm c1} = 0.67551 , 
\label{HN_44site_rc1}
\end{equation}
for $N_{\rm s}=44$ and 
\begin{equation}
r_{\rm c1} = 0.67542 ,
\label{HN_48site_rc1}
\end{equation}
for $N_{\rm s}=48$. 
One can recognize that the difference between Eqs.~(\ref{HN_44site_rc1}) and 
(\ref{HN_48site_rc1}) is quite small. 

\begin{figure}[tbh]
\includegraphics[width=15cm]{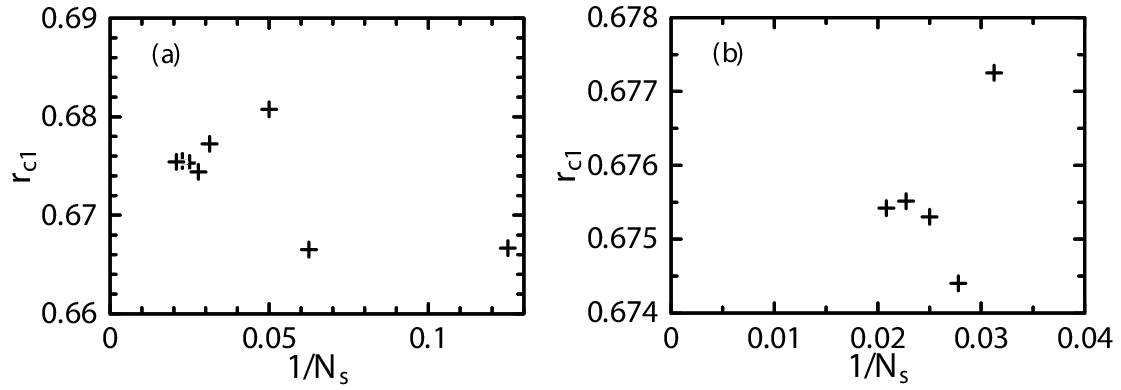}
\caption{System-size dependence 
of the ratio for the edge of the exact dimer phase. 
Panel (a) shows the entire range from $N_{\rm s}=8$ to 48. 
Panel (b) is a magnified view in the range from $N_{\rm s}=32$ to 48. 
}
\label{fig3}
\end{figure}

Next, let us examine the results of Eqs.~(\ref{HN_44site_rc1}) and 
(\ref{HN_48site_rc1}) together with those reported previously
in the studies based on the numerical-diagonalization method. 
The comparison is shown in Fig.~\ref{fig3}. 
When $N_{\rm s}$ decreases, 
$r_{\rm c1}$ shows a significantly large system-size dependence 
up to $N_{\rm s}=32$.  
On the other hand, the differences in estimated $r_{\rm c1}$ 
between neighboring data points become much smaller for $N_{\rm s}$
larger than 40. 
This behavior strongly suggests that 
our additional results for $N_{\rm s}=44$ and 48 
include only small finite-size deviations 
from the value in the thermodynamic limit. 
If the investigations treating even larger $N_{\rm s}$ 
based on the diagonalization method were carried out, 
it would be expected that 
differences between neighboring data points for $r_{\rm c1}$ 
become smaller. 
It is presently reasonable that an error is determined 
so that there are Eqs.~(\ref{HN_44site_rc1}) and 
(\ref{HN_48site_rc1}) within the error. 
Consequently, the final conclusion of this study is
\begin{equation}
r_{\rm c1}^{\rm (th)} = 0.6754(2) .
\label{HN_final_rc1}
\end{equation}

Let us discuss Eq.~(\ref{HN_final_rc1}) 
from the viewpoint of the comparison 
with the estimate in an experimental report. 
In Ref.~\ref{Sakurai_JPSJ2017}, 
Sakurai {\it et al}. carried out 
electron spin resonance measurements under high pressure and high field. 
They finally reported 
\begin{equation}
r_{\rm c1}^{\rm (ex)} = 0.660 \pm 0.003 .
\label{Sakurai_rc1}
\end{equation}
Then, one finds the difference to be
\begin{equation}
\left|r_{\rm c1}^{\rm (ex)}-r_{\rm c1}^{\rm (th)}\right|\sim 0.015 .
\label{difference1}
\end{equation}
The difference strongly suggests that 
the presence of other effects should be taken into account 
in addition to the simple Shastry-Sutherland Hamiltonian (\ref{Hamiltonian}). 
A high possibility is the presence of Dzyaloshinski-Moriya interactions. 
In Ref.~\ref{Sakurai_JPSJ2017}, 
Sakurai {\it et al}. also reported $J_1 = 69.1$~K,  $J_2/J_1=0.601$, and 
$D=1.6$~K at ambient pressure, 
where $D$ denotes the amplitude of the Dzyaloshinski-Moriya interaction 
along the $c$-axis 
on the bond forming the square lattice, 
namely, the bond of $J_2$;
therefore, the ratio of their estimates 
is found to be $D/J_1\sim 0.023$.
The relationship between this ratio and Eq.~(\ref{difference1}) 
should be examined in future studies together with other possibilities. 

\section{Summary and Remarks}

We have studied the $S=1/2$ Heisenberg antiferromagnet 
on the orthogonal-dimer lattice by the Lanczos diagonalization method. 
The boundary between the exact dimer phase and the neighboring phase 
is focused on; 
the corresponding ratio $J_2/J_1$ is found to be 0.6754(2).
Future studies should tackle estimations for other phase boundaries. 
One of them is the boundary between 
the N$\acute{\rm e}$el-ordered phase and an intermediate phase. 
Various estimates for this boundary in previous studies showed 
deviations that are more serious 
than the finite-size deviations of $r_{\rm c1}$. 
The other is the boundary inside the intermediate region
reported in Ref.~\ref{HNakano_JPSJ2018}. 
Such studies will greatly contribute 
to our fundamental understanding of magnetic materials with frustration. 

%\begin{acknowledgment}
%\acknowledgment
\section*{Acknowledgments}

%We wish to thank 
%Professors
%
%for fruitful discussions.
This work was partly supported 
by JSPS KAKENHI Grant Numbers 
16K05419, 16H01080(J-Physics), 18H04330(J-Physics), 
JP20K03866, and JP20H05274. 
%
%This work used computational resources
%of the K computer / the supercomputer Fugaku
%provided by RIKEN / ~~ provided by ››
%through the HPCI System Research Project (Project ID: hp######).
%
In this research, we used the computational resources 
of the supercomputer Fugaku provided by RIKEN 
through the HPCI System Research projects 
(Project IDs: hp200173, hp210068, hp210127, hp210201, and hp220043). 
Some of the computations were 
performed using facilities of 
the Institute for Solid State Physics, The University of Tokyo.

\end{document}